\documentclass{IET-Conf-Paper}
\usepackage{tikz}

\allowdisplaybreaks[4]

\theoremstyle{definition}

\newtheorem{remark}{Remark}
\newcommand{\qedblack}{\hfill \ensuremath{\blacksquare}}

\usepackage{balance}
\usepackage{multicol}

\newcommand{\tabincell}[2]{\begin{tabular}{@{}#1@{}}#2\end{tabular}}

\begin{document}

\title{A Unified Analytical Method to Quantify Three Types of Fast Frequency Response from Inverter-based Resources}

\author{~\\Shuan Dong\ad{1}, Xin Fang\ad{2}, Jin Tan\ad{1}\corr, Ningchao Gao\ad{1,3}, Xiaofan Cui\ad{4}, Anderson Hoke\ad{1}}

\address{\add{1}{Power Systems Energy Center, National Renewable Energy Laboratory, Golden, U.S.}
\add{2}{Department of Electrical and Computer Engineering, Mississippi State University, Starkville, U.S.}
\add{3}{Department of Electrical \& Computer Engineering, University of Denver, Denver, US}
\add{4}{Department of Energy Science \& Engineering, Stanford University, Stanford, U.S.}
\email{Jin.Tan@nrel.gov}}

\keywords{Fast frequency response (FFR), frequency nadir, inverter-based resources (IBRs), system frequency response model.}

\begin{abstract}
With more inverter-based resources (IBRs), our power systems have lower frequency nadirs following N-1 contingencies, and undesired under-frequency load shedding (UFLS) can occur. To address this challenge, IBRs can be programmed to provide at least three types of fast frequency response (FFR), e.g., step response, proportional response (P/f droop response), and derivative response (synthetic inertia). However, these heterogeneous FFR challenge the study of power system frequency dynamics. Thus, this paper develops an analytical frequency nadir prediction method that allows for the consideration of all three potential forms of FFR provided by IBRs. The proposed method provides fast and accurate frequency nadir estimation after N-1 generation tripping contingencies. Our method is grounded on the closed-form solution for the frequency nadir, which is solved from the second-order system frequency response model considering the governor dynamics and three types of FFR. The simulation results in the IEEE 39-bus system with different types of FFR demonstrate that the proposed method provides an accurate and fast prediction of the frequency nadir under various disturbances. 
\end{abstract}

\maketitle

%\balance

%%%%%%%%%%%%%%%%%%%%%%%%%%%%%%%%%%%%%%%%%%%%%%%%%%%%%%%%%%%%%%%%
\section{Introduction}
\begin{figure}
	\centering
	\begin{tikzpicture}[scale=5]
		\def\dy{0.3}
		% plot subfigure (a) 
		\draw[->,thick]  (0,0) -- (0,0.20)
		node[at end,left=31pt,rotate=90] {\small (a)~$f$ [Hz]};
		\draw[->,thick] (0,0) -- (1.20,0)
		node[at end, below=7pt, left] {\small Time [s]};
		\draw (0,0.15) node[anchor=east] {\small 60};
		%\draw[scale=1, line width=0.45mm, domain=0:1.2, smooth, variable=\x, black] plot ({\x}, {0.16-0.5*exp(-5*\x)*sin(2*3.14*50*\x-18});
		%\draw[scale=1, line width=0.45mm, domain=0:1.1, smooth, variable=\x, black] plot ({\x}, {0.12-0.6*exp(-5*\x)*sin(2*3.14*30*\x-13});
		\draw[line width=0.45mm, smooth] plot [smooth, tension=0.6] coordinates{(0,0.15) (0.25, 0.02) (0.6, 0.08) (1.1, 0.08)};
		\draw[dotted,line width=0.45mm] (0.28, 0.02) -- (0.28, 0.00);
		\draw (0.28, 0.00) node[anchor=north] {\small $t_{\rm nadir}$};
		\draw[dotted,line width=0.45mm] (0.28, 0.02) -- (0.00, 0.02);
		\draw (0.00, 0.02) node[anchor=east] {\small $f_{\rm nadir}$};
		\node at (0.28, 0.02) [circle,fill,inner sep=2pt]{};
		
		% plot subfigure (b)
		\draw[->,thick]  (0,0-\dy) -- (0,0.20-\dy)
		node[at end,left=31pt,rotate=90] {\small (b)~$P_{\it \rm ffr1}$};
		\draw[->,thick] (0,0-\dy) -- (1.20,0-\dy)
		node[at end, below=7pt, left] {\small Time [s]};  
		\draw[-,line width=0.45mm] (0.08,0-\dy) -- (0.20,0.15-\dy);
		\draw[-,line width=0.45mm] (0.20,0.15-\dy) -- (1.1,0.15-\dy);
		\draw[dotted,line width=0.45mm] (0.20,0.0-\dy) -- (0.20,0.15-\dy);
		\draw[dotted,line width=0.45mm] (0,0.15-\dy) -- (0.20,0.15-\dy);
		%  node[at end, below=10pt, left] {\small Time [s]};  
		\draw (0.08,0-\dy) node[anchor=north] {\small $t_1$};
		\draw (0.20,0-\dy) node[anchor=north] {\small $t_2$};
		\draw (0,0.15-\dy) node[anchor=east] {\small $P_{\rm sus}$};
		
		% plot subfigure (c)
		\draw[->,thick]  (0,0-2*\dy) -- (0,0.20-2*\dy)
		node[at end,left=31pt,rotate=90] {\small (c)~$P_{\it \rm ffr2}$};
		\draw[->,thick] (0,0-2*\dy) -- (1.20,0-2*\dy)
		node[at end, below=7pt, left] {\small Time [s]};  
		%\draw[scale=1, line width=0.45mm, domain=0:1.1, smooth, variable=\x, black] plot ({\x}, {0.16-2*\dy+0.7*exp(-5*\x)*sin(2*3.14*30*\x-13});
		\draw[line width=0.45mm, smooth] plot [smooth, tension=0.6] coordinates{(0,0-2*\dy) (0.25, 0.14-2*\dy) (0.6, 0.07-2*\dy) (1.1, 0.07-2*\dy)};
		
		% plot subfigure (d) 
		\draw[->,thick]  (0,0-3*\dy) -- (0,0.20-3*\dy)
		node[at end,left=31pt,rotate=90] {\small (d)~$P_{\it \rm ffr3}$};
		\draw[->,thick] (0,0-3*\dy) -- (1.20,0-3*\dy)
		node[at end, below=7pt, left] {\small Time [s]};  
		\draw[line width=0.45mm, smooth] plot [smooth, tension=0.5] coordinates{(0,-1.23+0.45-0.12) (0.04,-0.97+0.37-0.12) (0.25,-1.24+0.45-0.12) (0.5,-1.25+0.45-0.12) (0.7,-1.23+0.45-0.12) (1.1,-1.23+0.45-0.12)};
	\end{tikzpicture}
	\vspace{-5pt}
	\caption{Three main types of FFR provided by IBRs~\cite{nerc2020ffr}. (a)~grid frequency,~$f$, (b) step response,~$P_{\it \rm ffr1}$, (c) proportional response,~$P_{\it\rm ffr2}$, (d) derivative response,~$P_{\it\rm ffr3}$.}
	\label{fig:FFRs}
\end{figure}
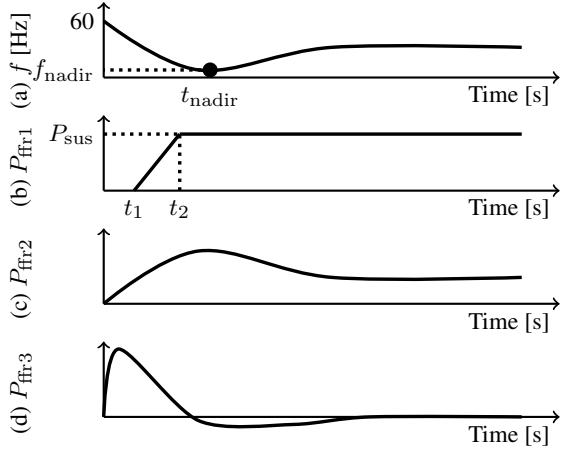

With the increasing share of inverter-based resources (IBRs), power systems are experiencing lower frequency nadirs following disturbances, such as generation trips and large renewable power variations \cite{Gao2022tpwrs}. Considering under-frequency load-shedding (UFLS) control, the lower frequency nadir might trip loads and cause a power outage. To tackle these challenges and keep the frequency nadir above the UFLS thresholds, one immediate solution is to leverage the capability of existing IBRs to provide fast frequency response (FFR). 
Here, FFR refers to the fast power injection in response to the frequency decline, aiming to increase the frequency nadir.
Based on~\cite{nerc2020ffr}, IBRs can provide three main types of FFR: step response (Fig.~\ref{fig:FFRs}(b)), proportional response (Fig.~\ref{fig:FFRs}(c)), and derivative~response~(Fig.~\ref{fig:FFRs}(d)).

With IBRs providing three types of FFR, one challenge for power system operators is how to quickly predict the frequency nadir after disturbances. This prediction method allows for determining whether power systems have enough FFR capacity to keep the frequency nadir above the UFLS threshold. Existing frequency nadir prediction methods mainly include simulation-based~\cite{doherty2005frequency} and analytical approaches~\cite{xiong2022performance,badesa2019simultaneous,liu2020analytical,guggilam2018optimizing}. Among them, the simulation-based method in \cite{doherty2005frequency} approximates the frequency nadir with a linear function; however, this requires creating a large database through numerous electromagnetic (EMT) simulations, which is known to be time-consuming and intractable for large-scale systems. Among analytical approaches, \cite{xiong2022performance} is computationally inexpensive, but omits the dynamic details of the turbine governors of the synchronous generator (SG) and the IBR in the system model. The methods in~\cite{badesa2019simultaneous} and~\cite{liu2020analytical} approximate the output of the SG turbine governor with a parameterized ramp function and a polynomial, respectively. They allow for the formulation of a tractable frequency-constrained optimization problem. But the accuracy of this frequency nadir approximation depends on the values of the selected parameters. The method in \cite{guggilam2018optimizing} considers a first-order turbine governor model and analytically characterizes frequency dynamics; however, it does not simultaneously consider the three types of FFR from IBRs. Indeed, to the best of our knowledge, few existing studies on fast frequency nadir prediction have fully considered the flexible combination of different types of FFR.

In this paper, we propose an accurate and efficient method to analytically predict the frequency nadir of power systems. The proposed method fully considers the impacts of three types of inverter-based FFR and provides a fast prediction speed, avoiding repetitive and time-consuming EMT simulation efforts.

%%%%%%%%%%%%%%%%%%%%%%%%%%%%%%%%%%%%%%%%%%%%%%%%%%%%%%%%%%%%%%%%
\section{Overview of Three Types of FFR from IBRs} \label{sec:ffr}
The heterogeneity in FFR designs within IBRs poses a serious challenge to the studies of power system frequency dynamics. To tackle this challenge, this section first overviews the three representative types of FFRs provided by IBRs (see Fig.~\ref{fig:FFRs}). As our first contribution, we develop the Laplace-domain models for these three types of FFRs to facilitate the system-level modeling and analysis of frequency dynamics. 

% \subsection{Three Major Types of FFR From IBRs}
We assume that the grid frequency drops as shown in Fig.~\ref{fig:FFRs}(a) following the N-1 contingency at $t=0~\mathrm{s}$. Then as illustrated in Figs.~\ref{fig:FFRs}(b)--(d), IBRs can provide three main types of FFR, i.e., step response $P_{\rm ffr1}$, proportional response $P_{\rm ffr2}$, and derivative response $P_{\rm ffr3}$, to contain the frequency deviation and avoid triggering UFLS. Below, we introduce these three types of FFR and propose their Laplace-domain models.

\begin{figure}[t!]
	\centering
	\includegraphics[width=1\linewidth]{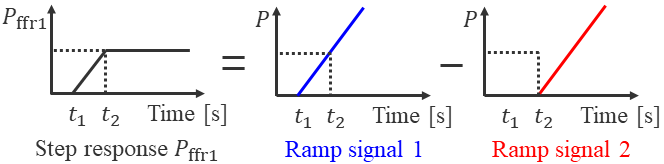}
	\caption{Decomposition of step response~$P_{\rm ffr1}$ into two ramp functions with time delays $t_1$ and $t_2$, respectively.} \label{fig:decouple}
\end{figure}

\textit{1) Step Response:} As shown in Fig.~\ref{fig:FFRs}(b), IBRs start to provide the step response after a time delay $t_1$ following the grid frequency drop at $t=0~\mathrm{s}$. Then the active power from IBRs ramps up until reaching the predefined saturation value~$P_{\rm sus}$ at $t=t_2$. Here, as displayed in Fig.~\ref{fig:decouple}, we notice that the step response $P_{\rm ffr1}$ in Fig.~\ref{fig:FFRs}(b) can be represented by the difference between two ramp functions with time delays~$t_1$ and~$t_2$, respectively. That is, we have the following time-domain decomposition for the step response
\begin{align}
	 P_{\it\rm ffr1} & = \underbrace{\dfrac{P_{\rm sus}(t-t_1)}{t_2-t_1} u(t-t_1)}_{\text{Ramp signal 1}} - \underbrace{\dfrac{P_{\rm sus}(t-t_2)}{t_2-t_1} u(t-t_2)}_{\text{Ramp signal 2}},   \label{eq:pffr1}
\end{align}
where~$u(t)$ denotes the unit step function.
Taking the Laplace transform of~\eqref{eq:pffr1}, we can obtain the Laplace-domain model of the step response as below:
\begin{align}
	\Delta P_{\rm ffr1}(s) = \frac{P_{\rm sus}}{(t_2 - t_1)s^2} e^{-t_1 s} - \frac{P_{\rm sus}}{(t_2 - t_1) s^2} e^{-t_2 s}. \label{eq:pffr1_s}
\end{align}
Note that in this paper, we let $\Delta(\cdot)$ denote the perturbations in variable~$(\cdot)$ after the disturbance.

\textit{2) Proportional Response:} As shown in Fig.~\ref{fig:FFRs}(c), IBRs can also provide proportional response~$P_{\rm ffr2}$ following a grid frequency drop. Note that the proportional response here is reminiscent of the P/f droop control or primary frequency control in that the IBR output change is proportional to the frequency deviation, as follows:
\begin{align}
	P_{\it\rm ffr2} = - \frac{1}{R_{\rm ibr}} (f - f_{\rm n}), \label{eq:pffr2}
\end{align}
where~$R_{\rm ibr}$ is the IBR droop coefficient, $f$ is the grid frequency, and $f_{\rm n}$ is the rated frequency. By taking the Laplace transform of~\eqref{eq:pffr2}, we obtain the Laplace-domain expression of the proportional response as follows:
\begin{align}
	\Delta P_{\it\rm ffr2} (s) = - \frac{1}{R_{\rm ibr}} \Delta f (s). \label{eq:pffr2_s}
\end{align}
Note that we assume~$f_{\rm n}$ remain unchanged, and thus $\Delta f_{\rm n} = 0$.

\textit{3) Derivative Response:} The derivative response~$P_{\rm ffr3}$ in Fig.~\ref{fig:FFRs}(d) allows IBRs to provide "synthetic inertia" to power systems. To achieve this, $P_{\rm ffr3}$ is controlled to be proportional to the time derivative of the measured grid frequency, as follows:
\begin{align}
	P_{\it\rm ffr3} = - 2 H_{\rm ibr} \frac{df}{dt},  \label{eq:pffr3}
\end{align}
in which $H_{\rm ibr}$ denotes the IBR emulated inertia. 
In the Laplace domain, the derivative response can be expressed as
\begin{align}
	\Delta P_{\it\rm ffr3}(s) = - 2 H_{\rm ibr} s \Delta f(s). \label{eq:pffr3_s}
\end{align}
\begin{remark} Here, we neglect the dynamics of IBR power controllers, whether grid following (GFL) or grid forming (GFM) controllers, as they are much faster than conventional SG and governor dynamics. In addition, we do not differentiate GFL and GFM in this paper, because both can provide different types of FFR or a combination of them. \qedblack
\end{remark}

%%%%%%%%%%%%%%%%%%%%%%%%%%%%%%%%%%%%%%%%%%%%%%%%%%%%%%%%%%%%%%%%
\section{Improved System Frequency Response Model}

In this section, we first improve the conventional system frequency response (SFR) model by including the three types of FFR in Section~\ref{sec:ffr}. Our improved second-order SFR model can accurately predict the frequency dynamics of power systems with high penetration of IBRs. Thereafter, by solving this second-order SFR model, we obtain the analytical expression of the post-disturbance system frequency. 
This expression of frequency nadir allows us to directly predict the post-disturbance frequency nadir without onerous simulation. In this way, we can easily see whether the current FFR settings avoid triggering the ULFS under a given disturbance. 

\subsection{Including FFR Into System Frequency Response Model}

\begin{figure}[t!]
	\centering
	\includegraphics[width=1\linewidth]{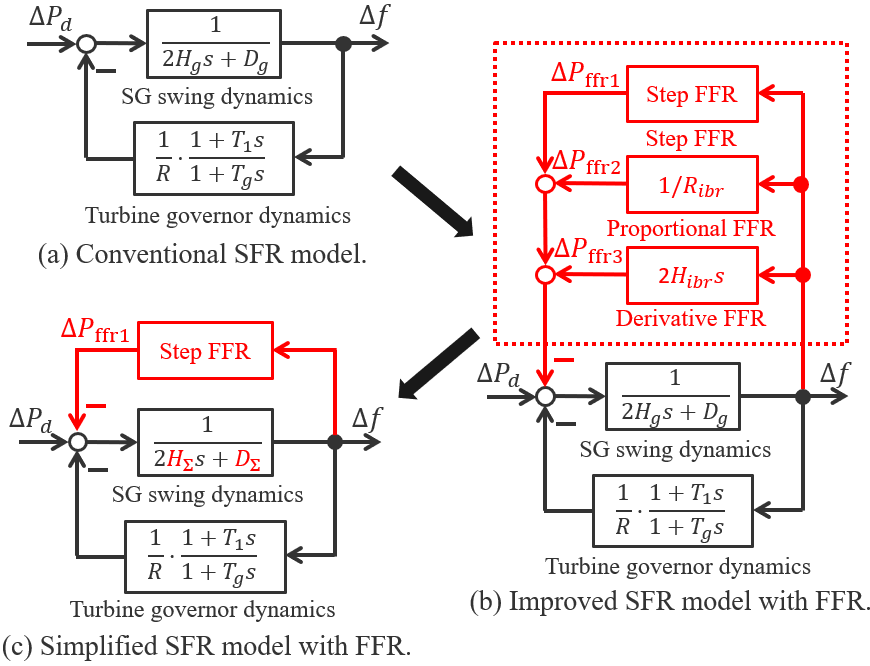}
	\caption{SFR models. (a)~Conventional SFR model. (b)~Improved SFR model with three types of FFR from IBRs. (c)~Simplified SFR model with three types of FFR from IBRs. Note that we combine proportional FFR and derivative FFR into the SG damping and inertia constants, respectively (by defining~$H_\Sigma:=H_{\rm g} + H_{\rm ibr}$ and $D_\Sigma:=D_{\rm g} + R_{\rm ibr}^{-1}$).} \label{fig:SFR}
\end{figure}

The conventional system frequency response (SFR) model~\cite{anderson1990low} enables us to analytically predict the frequency dynamics and evaluate the impacts of different key parameters. However, the conventional SFR model is developed for SG-dominated power systems, and thus it has not considered the three main types of FFR provided by IBRs. As our second contribution, we will include all three types of FFR in the SFR model while ensuring that our proposed SFR model remains analytically tractable.

The conventional SFR model in Fig.~\ref{fig:SFR}(a) assumes uniform frequency dynamics, omits the impact of the network, and only considers the dynamics of SG rotors and governors. Therefore, we propose an improved SFR model with three types of FFR (highlighted in red) as shown in Fig.~\ref{fig:SFR}(c). We obtain this improved SFR model as follows. First, in the Laplace domain, the system frequency dynamics are delineated by the swing equation~\eqref{eq:swing1} and the governor~\eqref{eq:gov} as below:
\begin{align}
	2 H_{\rm g} s \Delta f =&~\Delta P_{\rm m} \underbrace{- \Delta P_{\rm load}}_{=: \Delta P_d} + \underbrace{\Delta P_{\it\rm ffr1} +\Delta P_{\it\rm ffr2} + \Delta P_{\it\rm ffr3}}_{\text{Three types of FFR}} \nonumber \\ 
	&\underbrace{- D_{\rm g}(\Delta f - \Delta f_{\rm n})}_{\text{Damping}}, \label{eq:swing1} \\
	\Delta P_m =&~ \underbrace{\frac{1+T_1 s}{1+T_g s} \cdot \left( \Delta P_{\rm m}^\star - \frac{1}{R_{\rm g}}(\Delta f- \Delta f_{\rm n}) \right)}_{\text{Turbine governor output}}. \label{eq:gov}
\end{align}
Note that in the SG swing dynamics, we have included three types of FFR provided by IBR. In~\eqref{eq:swing1} and~\eqref{eq:gov}, $H_{\rm g}$ is the SG inertia, $P_{\rm m}$ and $P_{\rm m}^\star$ are the output of the SG turbine governor and its reference, $P_{\rm load}$ is the load, $D_{\rm g}$ is the damping constant, $T_1$ and~$T_{\rm g}$ are the SG governor time constants, and $R_g$ is the SG droop coefficient. Note that we define the system disturbance as $\Delta P_d := - \Delta P_{\rm load}$. Then we assume that the rated frequency~$f_n$ and the turbine governor reference~$P_m^\star$ remain unchanged, i.e., $\Delta f_n = 0$ and~$\Delta P_m^\star = 0$. By substituting~\eqref{eq:pffr2_s} and~\eqref{eq:pffr3_s} into~\eqref{eq:swing1} and~\eqref{eq:gov}, we have the simplified SFR model as below.
\begin{align}
	2 \underbrace{(H_{\rm g}\! +\! H_{\rm ibr})}_{=: H_{\Sigma}} s \Delta f =&~\Delta P_{\rm m} \! +\! \Delta P_d \! +\! \Delta P_{\it\rm ffr1} \! -\! \underbrace{(D_{\rm g} \! +\! R_{\rm ibr}^{-1})}_{=: D_{\Sigma}} \Delta f ,  \label{eq:swing2} \\
	\Delta P_m =&~ - \frac{1}{R_{\rm g}} \cdot \frac{1+T_1 s}{1+T_g s} \cdot \Delta f. \label{eq:gov2}
\end{align}
We note that by defining~$D_\Sigma:=D_g + R_{\rm ibr}^{-1}$ and~$H_\Sigma := H_{\rm g}+ H_{\rm ibr}$, the IBR droop coefficient~$R_{\rm ibr}$ and the IBR emulated inertia~$H_{\rm ibr}$ are, respectively, absorbed into the total damping~$D_\Sigma$ and the total inertia~$H_\Sigma$. This is clearer when we visualize the simplified SFR model~\eqref{eq:swing2} and~\eqref{eq:gov2} in Fig.~\ref{fig:SFR}(c). 

We also highlight that our final simplified SFR model not only considers three types of FFR but also remains analytically tractable since it is still a second-order model. To show this, we treat the term~$(\Delta P_{\rm ffr1} + \Delta P_d)$ as input and get the following Laplace-domain expression of the frequency deviation~$\Delta f(s)$:
\begin{align}
	\Delta f(s) =  \frac{s + T_{\rm g}^{-1} }{s^2 + 2 \zeta \omega_{\rm n}s + \omega_{\rm n}^2} \cdot \frac{\Delta P_{\it\rm ffr1}(s) + \Delta P_{\rm d}(s)}{2 H_\Sigma}, \label{eq:sys3}
\end{align}
where the damping ratio~$\zeta$, natural frequency~$\omega_{\rm n}$, and damped natural frequency~$\omega_{\rm d}$ satisfy:
\begin{align}
	\zeta &= \dfrac{1}{2} \left( \dfrac{1}{2 H_\Sigma} \left( D_\Sigma + \frac{T_1}{T_g} R_g^{-1} \right) + \dfrac{1}{T_{\rm g}} \right) \sqrt{\dfrac{{2 T_{\rm g} H_\Sigma}}{D_\Sigma + R_{\rm g}^{-1}}} \nonumber \\ &=: \cos{\phi},  \label{eq:zeta} \\
	\omega_{\rm n} &= \sqrt{\dfrac{D_\Sigma + R_{\rm g}^{-1}}{2 T_{\rm g} H_\Sigma}} =: \frac{\omega_{\rm d}}{\sqrt{1-\zeta^2}}. \label{eq:wn}
\end{align}

%%%%%%%%%%%%%%%%%%%%%%%%%%%%%%%%%%%%%%%%%%%%%%%%%%%%%%%%%%%%%%%%
\subsection{Proposed Frequency Nadir Computation Method} \label{sec:method}
With the second-order SFR model in Fig.~\ref{fig:SFR}(c) and \eqref{eq:sys3}--\eqref{eq:wn}, this section develops the analytical expression of the frequency nadir~$f_{\rm nadir}$ in this section. Our expression of frequency nadir will allow us to directly predict the post-disturbance frequency nadir without onerous simulation. In this way, we can easily see whether the current FFR settings avoid triggering the ULFS under a given disturbance.

\begin{figure*}[t!] %\footnotesize	
	\begin{align}
		\Delta f(t) &= \dfrac{P_{\rm sus} + \Delta P_{\rm d}}{D_\Sigma + R_{\rm g}^{-1}} + \frac{\sqrt{\vphantom{M^2} \omega_{\rm n}^2 - 2 \zeta T_{\rm g}^{-1} + T_{\rm g}^{-2}}}{2 H_\Sigma  \omega_{\rm n}^2 \omega_{\rm d} (t_2-t_1)} \cdot e^{-\zeta \omega_{\rm n} t} M \sin{\left( \omega_{\rm d} t + \alpha  \right)}, ~~~~~~~~~~~~t>t_2, \tag{$\dagger$}\label{eq:ft}   \\
		\text{where}~M &= \sqrt{\left( m(0) \right)^2 + \left( m\left(\dfrac{\pi}{2 \omega_{\rm d}}\right) \right)^2}, ~~\alpha = \pi - \sin^{-1}{\left( \frac{m(0)}{M} \right)}, ~~\beta = \sin^{-1}{\left( \dfrac{T_{\rm g}^{-1} \sqrt{\vphantom{M^2} 1 - \zeta^2}}{\sqrt{ \vphantom{M^2} \omega_{\rm n}^2 - 2 \zeta T_{\rm g}^{-1} + T_{\rm g}^{-2}}} \right)},  \nonumber \\
		% m(t) &= -e^{\zeta \omega_{\rm n} t_1} \sin{\left( \omega_{\rm d} (t-t_1) - \beta - \phi \right)} + e^{\zeta \omega_{\rm n} t_2} \sin{\left( \omega_{\rm d} (t-t_2) - \beta - \phi \right)} - \dfrac{\Delta P_{\rm load}}{P_{\rm sus}} \omega_{\rm n} (t_2-t_1) \sin{\left( \omega_{\rm d} t - \beta \right)}. \nonumber \\
		m(t) &= P_{\rm sus} e^{\zeta \omega_{\rm n} t_2} \sin{\left( \omega_{\rm d} (t-t_2)- \beta - \phi \right)} -P_{\rm sus} e^{\zeta \omega_{\rm n} t_1} \sin{\left( \omega_{\rm d} (t-t_1) - \beta - \phi \right)}  + \Delta P_{\rm d} \, \omega_{\rm n} (t_2-t_1) \sin{\left( \omega_{\rm d} t - \beta \right)}. \nonumber
	\end{align}
	\hrulefill 
\end{figure*}

To begin with, we recall that the input of our developed FFR model~\eqref{eq:sys3}--\eqref{eq:wn} is
\begin{align}
	\Delta P_{\it\rm ffr1}(s) + \Delta P_{\rm d} = \dfrac{P_{\rm sus} \left( e^{-s t_1} - e^{-s t_2} \right)}{(t_2 - t_1) s^2} + \frac{\Delta P_{\rm d}}{s}. \label{eq:input}
\end{align}
Substituting~\eqref{eq:input} into~\eqref{eq:sys3} and taking the inverse Laplace transform, we obtain the closed-form time-domain expression of the post-disturbance frequency~$\Delta f(t)$ as in~\eqref{eq:ft}. In~\eqref{eq:ft}, we consider only the frequency dynamics for~$t>t_2$ because the frequency nadir time~$t_{\rm nadir}$ in Fig.~\ref{fig:FFRs}(a) is typically larger than the time~$t_2$ of fully implementing the step response in Fig.~\ref{fig:FFRs}(b).
Then, by taking the time derivative of~$\Delta f(t)$ in~\eqref{eq:ft} and solving~$\frac{d\Delta f(t)}{dt}=0$, we get the frequency nadir time:
\begin{align}
	t_{\rm nadir} = \frac{\pi + \phi - \alpha}{\omega_{\rm d}}. \label{eq:tnadir}
\end{align}
Finally, setting~$t=t_{\rm nadir}$ in~\eqref{eq:ft} yields the following analytical equation, which allows us to directly compute the frequency nadir:
\begin{align}
	f_{\rm nadir} \!=\! f_{\rm n} + \dfrac{P_{\rm sus} + \Delta P_{\rm d}}{D_\Sigma + R_{\rm g}^{-1}} - \dfrac{T_{\rm g} R_{\rm g}^{-\frac{1}{2}} M e^{(\alpha-\phi-\pi) \cot{\phi}}}{(t_2-t_1) \left( D_\Sigma  + R_{\rm g}^{-1} \right)^{\frac{3}{2}}}. \label{eq:fnadir}
\end{align}

\begin{figure}[t!]
	\centering
	\includegraphics[width=1\linewidth]{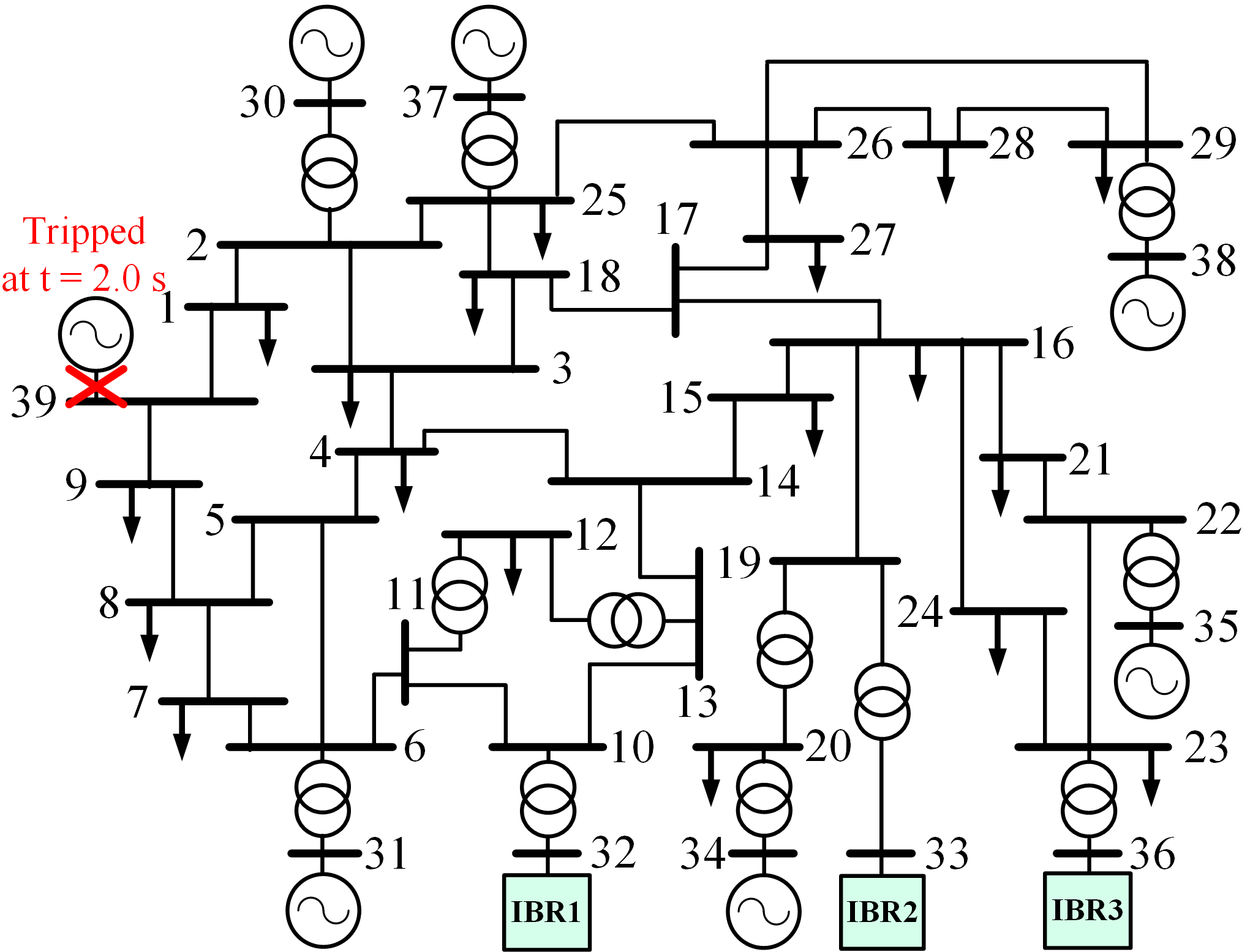}
	
%	\vspace{-6pt}
	\caption{Modified IEEE 39-bus test system used to validate the proposed frequency nadir prediction method.} \label{fig:grid}
\end{figure}

\begin{figure}[t!]
	\centering
	{\includegraphics[width=1\linewidth]{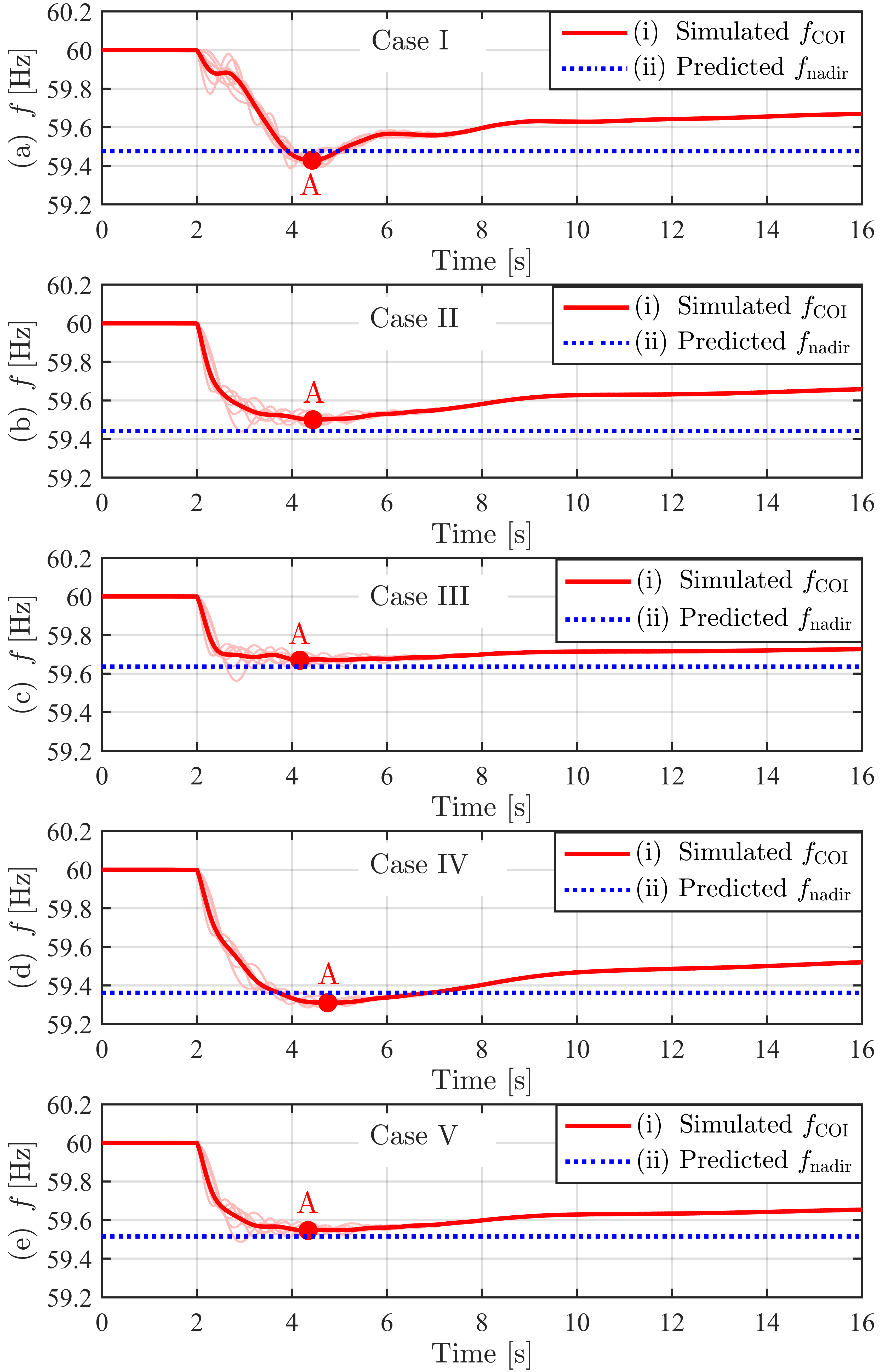}}
	
	\vspace{-5pt}
	\caption{Comparison of simulated and predicted frequency nadirs in cases~I--V.} \label{fig:simu}
\end{figure}

In sum, our proposed frequency nadir prediction method is as follows. First, we describe the system frequency dynamics with the conventional system frequency response model in~Fig.~\ref{fig:SFR}(b). This step can be achieved by referring to~\cite{guggilam2018optimizing, anderson1990low,shi2018analytical}. Next, we represent the proportional response~$P_{\it\rm ffr2}$ and the derivative response~$P_{\it\rm ffr3}$ with the equivalent damping and inertia constants as shown in Fig.~\ref{fig:SFR}(c). In this way, we revise the system frequency response model to be~\eqref{eq:swing2} and~\eqref{eq:gov2}. Last, under the given disturbance~$\Delta P_{\rm d}$, we compute the resultant frequency nadir~$f_{\rm nadir}$ directly with~\eqref{eq:fnadir}.

%%%%%%%%%%%%%%%%%%%%%%%%%%%%%%%%%%%%%%%%%%%%%%%%%%%%%%%%%%%%%%%%%%%%%%%%%%%%
\section{Simulation Results} 

This section validates both the accuracy and the efficiency of our proposed frequency nadir prediction method with the modified 39-bus test system, as shown in Fig.~\ref{fig:grid}. Note that the rated capacities of the SGs or IBRs connected to buses~38 and~39 and buses~31--37 are, respectively, $1500$ and~$1000~\mathrm{MVA}$. All SGs are equipped with DC1A exciters and the IEEEG1 steam governor model, with the parameters reported in Table~\ref{tab:para} and~\cite{sauer1998power}. The IBRs can represent inverter-interfaced batteries or hybrid photovoltaic (PV) plants that combine PV and batteries. We assume that all IBRs adopt conventional GFL controllers.

\begin{table}[t!] \footnotesize 
	\caption{\label{tab:para} Parameters of SG IEEEG1 steam governor and three FFR} 
	
%	\vspace{-6pt}
	\begin{tabular}{ c|c|c|c|c|c|c|c|c } 
		DB & $R_g$ & TSR & TSM & $K_1$ & $K_2$, $K_3$ & $K_4$--$K_8$ & $T_4$ &  $T_5$  \\ \hline
		0 & 0.05 & 0 & 0.075 & 0.2 & 0.4 & 0 & 0.3 & 10 
	\end{tabular}
	
	\vspace{6pt}
	\begin{tabular}{ c|c||c|c|c||c||c} 
		$T_6$ & $T_7$ & $t_1$\,[s] & $t_2$\,[s] & $P_{\rm sus}$\,[MW] & $R_{\rm ibr}$\,[p.u.] & $H_{\rm ibr}$\,[s] \\ \hline
		0.6 & 0 & 0.05 & 0.35 & 100 & 0.03 & 4 
	\end{tabular}
\end{table}

\emph{1)~Verification of Prediction Accuracy:} We consider cases~I--V (summarized in Table~\ref{tab:veri1}) to show the accuracy of the proposed prediction method by comparing the simulated and predicted frequency nadir~$f_{\rm nadir}$. In case~I, we keep the SGs at buses~32, 33, and~36, whereas in cases~II--V,  we replace these three SGs with IBRs providing different combinations of FFR types (see Table~\ref{tab:para} for FFR parameters). With these settings in place, we simulate the 39-bus system in PSCAD/EMTDC and trip the SG connected to Bus~39 at~$t=2~\mathrm{s}$ to get the actual~$f_{\rm nadir}$ of the center-of-inertia frequency~$f_{\rm COI}$. Also, we predict~$f_{\rm nadir}$ with our proposed method in Section~\ref{sec:method}. As shown in Fig.~\ref{fig:simu}, the predicted~$f_{\rm nadir}$ (blue dotted trace) is close to the simulated one (red point A) in cases~I--V. As reported in Table~\ref{tab:veri1}, all prediction errors in cases~I--V are within~$0.06~\mathrm{Hz}$.

\emph{2)~Verification of Prediction Efficiency:} We further leverage cases~V--XI to show the efficiency of the proposed prediction method. As reported in Table~\ref{tab:veri2}, our verification is achieved by comparing the simulation and prediction time in each case. Note that in cases~V--XI, we have the same FFR settings but trip different SGs. Taking case~XI in Table~\ref{tab:veri2} as an example, when tripping the SG at Bus 38, we need 988.55 s to obtain~$f_{\rm nadir}$ via simulation in PSCAD/EMTDC, but our proposed method takes only~$0.15~\mathrm{ms}$, and the prediction error is within~$0.06~\mathrm{Hz}$. This is also true for other cases in Table~\ref{tab:veri2}; thus, our prediction method is~$10^6$~times faster than the simulation-based method.

\begin{table}[t!] \footnotesize 
	\caption{\label{tab:veri1} Accuracy of proposed frequency nadir prediction method} 
	
	%	\vspace{-6pt}
	\begin{tabular}{ >{\centering}p{0.5cm}|c|c|c|c|c|c } 
		Case & Bus32 & Bus33 & Bus36 & \tabincell{c}{Simulated \\ $f_{\rm nadir}$} & \tabincell{c}{Predicted \\ $f_{\rm nadir}$} & \tabincell{c}{Prediction \\ error} \\ \hline
		I & SG & SG & SG & 59.43 Hz & 59.48 Hz & 0.05 Hz \\ 
		II & $P_{\it\rm ffr1}$ & $P_{\it\rm ffr1}$ & $P_{\it\rm ffr1}$ & 59.50 Hz & 59.44 Hz & 0.06 Hz \\
		III & $P_{\it\rm ffr2}$ & $P_{\it\rm ffr2}$ & $P_{\it\rm ffr2}$ & 59.67 Hz & 59.64 Hz & 0.03 Hz \\
		IV & $P_{\it\rm ffr3}$ & $P_{\it\rm ffr3}$ & $P_{\it\rm ffr3}$ & 59.31 Hz & 59.36 Hz & 0.05 Hz \\
		V & $P_{\it\rm ffr2}$ & $P_{\it\rm ffr3}$ & $P_{\it\rm ffr1}$ & 59.55 Hz & 59.52 Hz & 0.03 Hz 
	\end{tabular}
\end{table}

%%%%%%%%%%%%%%%%%%%%%%%%%%%%%%%%%%%%%%%%%%%%%%%%%%%%%%%%%%%%%%%%%%%%%%%%%%%%%
\section{Concluding Remarks} 
This paper proposes an analytical method to predict the frequency nadir of power systems with three types of FFR provided by IBRs. The nonlinear dynamics of the turbine governor's response and the IBR FFR following generation trip events are captured in our derivation of the nadir formulation. Our prediction method not only demonstrates high prediction accuracy but also exhibits extra-fast prediction speed compared to traditional time-domain simulations. 

% Compelling future directions include the frequency nadir constraint formulation in generation scheduling optimization problems, such as evaluation of FFR capacity adequacy from IBRs, unit commitment and economic dispatch, the application of the prediction method in power system real-time security monitoring, and the adaptation of our method for wide-area power systems.

Compelling future directions include frequency nadir constraint formulation in generation scheduling optimization problems (e.g., unit commitment and economic dispatch), evaluation of FFR capacity adequacy from IBRs, and the application of the prediction method in real-time power system security monitoring.

\begin{table}[t!] \footnotesize 
	\caption{\label{tab:veri2} Efficiency of proposed frequency nadir prediction method} 
	
	%	\vspace{-6pt}
	\begin{tabular}{>{\centering}p{0.5cm}|c|c|c|c|c} 
		Case & \tabincell{c}{Tripped  \\ SG} & \tabincell{c}{Simulation \\ time} &  \tabincell{c}{Prediction \\ time} & Speed-up & \tabincell{c}{Prediction \\ error} \\ \hline
		V & Bus39 & 1087.13 s & 0.12 ms & $8.97\!\times\!10^6$ & 0.03 Hz \\
		VI & Bus30 & 1026.94 s & 0.12 ms & $8.49\!\times\!10^6$ & 0.02 Hz \\
		VII & Bus31 & 1063.95 s & 0.18 ms & $6.02\!\times\!10^6$ & 0.06 Hz \\
		VIII & Bus34 & 1009.02 s & 0.15 ms & $6.81\!\times\!10^6$ & 0.05 Hz \\
		IX & Bus35 & ~904.28 s & 0.15 ms & $5.84\!\times\!10^6$ & 0.05 Hz \\
		X & Bus37 & ~928.84 s & 0.15 ms & $6.26\!\times\!10^6$ & 0.03 Hz  \\
		XI & Bus38 & ~988.55 s & 0.15 ms & $6.62\!\times\!10^6$ & 0.06 Hz
	\end{tabular}
\end{table}

%%%%%%%%%%%%%%%%%%%%%%%%%%%%%%%%%%%%%%%%%%%%%%%%%%%%%%%%%%%%%%%%
\section{Acknowledgements}
This work was authored in part by the National Renewable Energy Laboratory, operated by Alliance for Sustainable Energy, LLC, for the U.S. Department of Energy (DOE) under Contract No. DE-AC36-08GO28308. This material is based upon work supported by the U.S. Department of Energy's Office of Energy Efficiency and Renewable Energy (EERE) under the Solar Energy Technologies Office Award Number 37772. The U.S. Government retains and the publisher, by accepting the article for publication, acknowledges that the U.S. Government retains a nonexclusive, paid-up, irrevocable, worldwide license to publish or reproduce the published form of this work, or allow others to do so, for U.S. Government purposes. The views expressed herein do not necessarily represent the views of the U.S. Department of Energy or the United States Government.

%%%%%%%%%%%%%%%%%%%%%%%%%%%%%%%%%%%%%%%%%%%%%%%%%%%%%%%%%%%%%%%%

\end{document}